\documentclass[aps,
twocolumn,
showpacs,preprintnumbers,amsmath,amssymb,prb,superscriptaddress]{revtex4-1}
\usepackage{graphicx}
\usepackage{amssymb}
\usepackage{amsmath}
\usepackage{enumitem}
\usepackage{color}
\usepackage{textcomp}
\usepackage[symbol]{footmisc}

\begin{document}

\title{Enhancing fluorescence excitation and collection from the nitrogen-vacancy center in diamond through a micro-concave mirror}

\author{Dewen Duan}
\email{dduan@gwdg.de}
\affiliation{Max-Planck Research Group Nanoscale Spin Imaging, Max Planck Institute for Biophysical Chemistry,Am Fassberg 11,G\"{o}ttingen, 37077, Germany}
\author{Vinaya Kumar Kavatamane}
\affiliation{Max-Planck Research Group Nanoscale Spin Imaging, Max Planck Institute for Biophysical Chemistry,Am Fassberg 11,G\"{o}ttingen, 37077, Germany}
\author{Sri Ranjini Arumugam}
\affiliation{Max-Planck Research Group Nanoscale Spin Imaging, Max Planck Institute for Biophysical Chemistry,Am Fassberg 11,G\"{o}ttingen, 37077, Germany}
\author{Ganesh Rahane}
\affiliation{Max-Planck Research Group Nanoscale Spin Imaging, Max Planck Institute for Biophysical Chemistry,Am Fassberg 11,G\"{o}ttingen, 37077, Germany}
\author{Yan-Kai Tzeng}
\affiliation{Institute of Atomic and Molecular Sciences, Academia Sinica, Taipei 106, Taiwan}
\affiliation{Current adress: Department of Physics, Stanford University, Stanford, California 94305, USA}
\author{Huan-Cheng Chang}
\affiliation{Institute of Atomic and Molecular Sciences, Academia Sinica, Taipei 106, Taiwan}
\author{Hitoshi Sumiya}
\affiliation{Sumitomo Electric Industries Ltd., Itami 664-001, Japan}
\author{Shinobu Onoda}
\affiliation{Takasaki Advanced Radiation Research Institute, National Institutes for Quantum and Radiological Science and Technology, Takasaki, Gunma 370-1292, Japan}
\author{Junichi Isoya}
\affiliation{Research Center for Knowledge Communities, University of Tsukuba, 1-2 Kasuga, Tsukuba, Ibaraki 305-8550, Japan}
\author{Gopalakrishnan Balasubramanian}
\email{gbalasu@gwdg.de}
\affiliation{Max-Planck Research Group Nanoscale Spin Imaging, Max Planck Institute for Biophysical Chemistry,Am Fassberg 11,G\"{o}ttingen, 37077, Germany}
\affiliation{Center Nanoscale Microscopy and Molecular Physiology of the Brain (CNMPB), G\"{o}ttingen, 37077, Germany}

\begin{abstract}

We experimentally demonstrate a simple and robust optical fibers based method to achieve simultaneously efficient excitation and fluorescence collection from Nitrogen-Vacancy (NV) defects containing micro-crystalline diamond. We fabricate a suitable micro-concave (MC)  mirror that focuses scattered excitation laser light into the diamond located at the focal point of the mirror. At the same instance, the mirror also couples the fluorescence light exiting out of the diamond crystal in the opposite direction of the optical fiber back into the optical fiber within its light acceptance cone. This part of fluorescence would have been otherwise lost from reaching the detector. Our proof-of-principle demonstration achieves a 25 times improvement in fluorescence collection compared to the case of not using any mirrors. The increase in light collection favors getting high signal-to-noise ratio (SNR) optically detected magnetic resonance (ODMR) signals hence offers a practical advantage in fiber-based NV quantum sensors. Additionally, we compacted the NV sensor system by replacing some bulky optical elements in the optical path with a $ 1\times2 $ fiber optical coupler in our optical system. This reduces the complexity of the system and provides portability and robustness needed for applications like magnetic endoscopy and remote-magnetic sensing. \\

\end{abstract}

\maketitle

Negatively charged Nitrogen-Vacancy (NV) color center in diamond is a promising candidate for quantum information processing and spin-based quantum sensing of magnetic fields\cite{Chipaux2015, Fedotov2016, Glenn2018}, electric fields\cite{Dolde2011, Dolde2014} and temperatures\cite{Acosta2010, Kucsko2013, Neumann2013}. To efficiently use NV center in these applications, enhancing the excitation and collection of the color-center's fluorescence is demanded. For instance a high fluorescence collection efficiency would benefit precision metrology. Some methods; for example, circular gratings \cite{Li2015, Andersen2018} or a solid immersion lenses (SIL)\cite{Hadden2010, Marseglia2011, Jamali2014}, when fabricated on diamond substrates, extracts more fluorescence from a single NV center out of the high refractive index ($\sim$2.4) material. While Micro/nanoresonator cavities and tapered optical fiber have been shown to enhance both the single NV center's fluorescence emission and coupling-out efficiencies\cite{Babinec2011, Liebermeister2014, Janitz2015}. Similarly, optical antennas and plasmonic structures have been used to enhance the local light field thus improving the fluorescence efficiency through the Purcell effect, and enhance emission from a single NV defect in diamond\cite{ Barth2010, Schroder2016}.   

Here, we demonstrate a simple method to enhance the excitation efficiency and also fluorescence collection from a micrometer-sized NV color-center rich diamond attached to the end of a sphered optical fiber. This single-port sensor configuration offers flexibility needed for some applications that require precision field-measurements in a non-standard location (e.g. magnetic endoscopy, subcutaneous measurements, remote and inaccessible locations). Our enhancement technique is based on using a matched micro-concave (MC) mirror to a sphered optic-fiber end. 

A diamond micro-crystal is fixed on the sphered tip optical fiber and excited by 532 nm laser light guided through the fiber. The diamond scatters  incident light except for a fraction that is absorbed by the NV defects. The MC mirror reflects and focuses the scattered laser light into the diamond (on the sphered fiber tip) that is positioned at the focal point of the MC mirror. At the same time, the MC mirror also serves a dual purpose to reflects and focuses the fluorescence light that is scattered out of the diamond into the optical fiber within its acceptance angle thereby increasing the effective numerical aperture of the light collection system. In a proof-of-principle demonstration and validation experiment, we were able to collect over 25 times higher fluorescence intensity from the diamond located at the focal point of an MC mirror compared to the intensity collected without any mirrors.  Although we only demonstrated the MC mirror for enhancing excitation and fluorescence collection of NV ensembles in micrometer-sized diamond. We could foresee this method can improve laser excitation and fluorescence collection of single NV centers and other solid-state quantum emitters such as Silicon-Vacancy centers in diamond, silicon-vacancies in Silicon carbide and rare-earth ions. In addition, we simplified the optical fiber-based NV sensor system by using a fiber optical coupler to replace the bulk lens and dichroic-beam-splitters based optical path splitter. This replacement reduces the complexity, enhance the portability, and improve the robustness of this optical-fiber-based NV system.

\begin{figure}[htbp]
\centering
\fbox{\includegraphics[width=0.95\columnwidth]{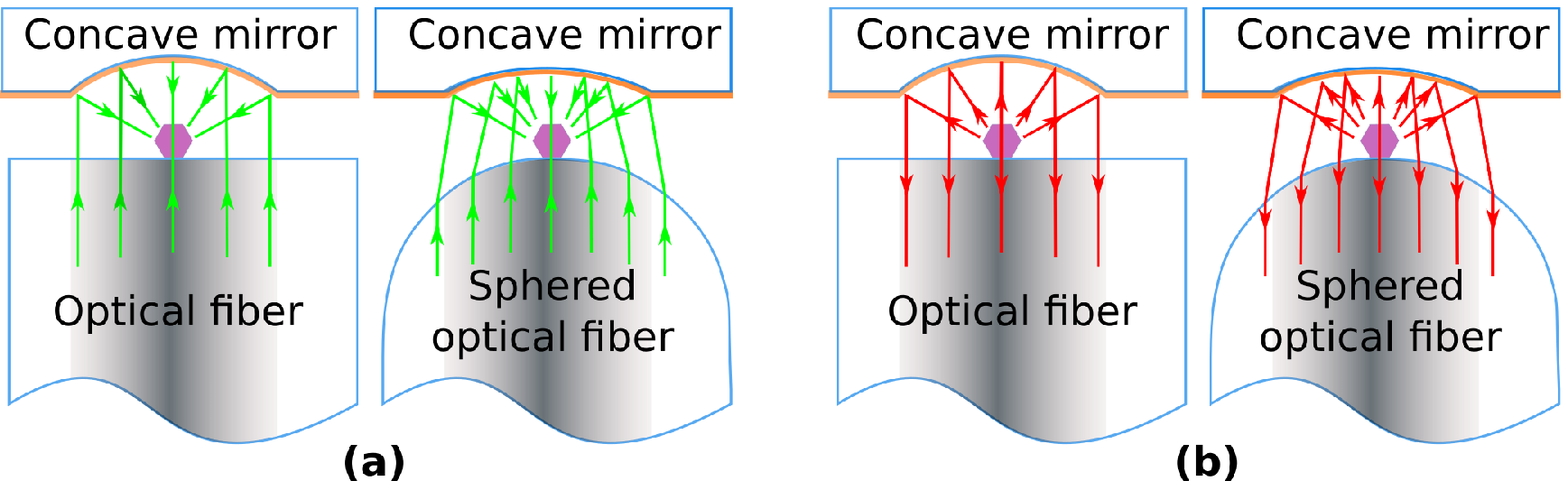}}
\fbox{\includegraphics[width=0.95\columnwidth]{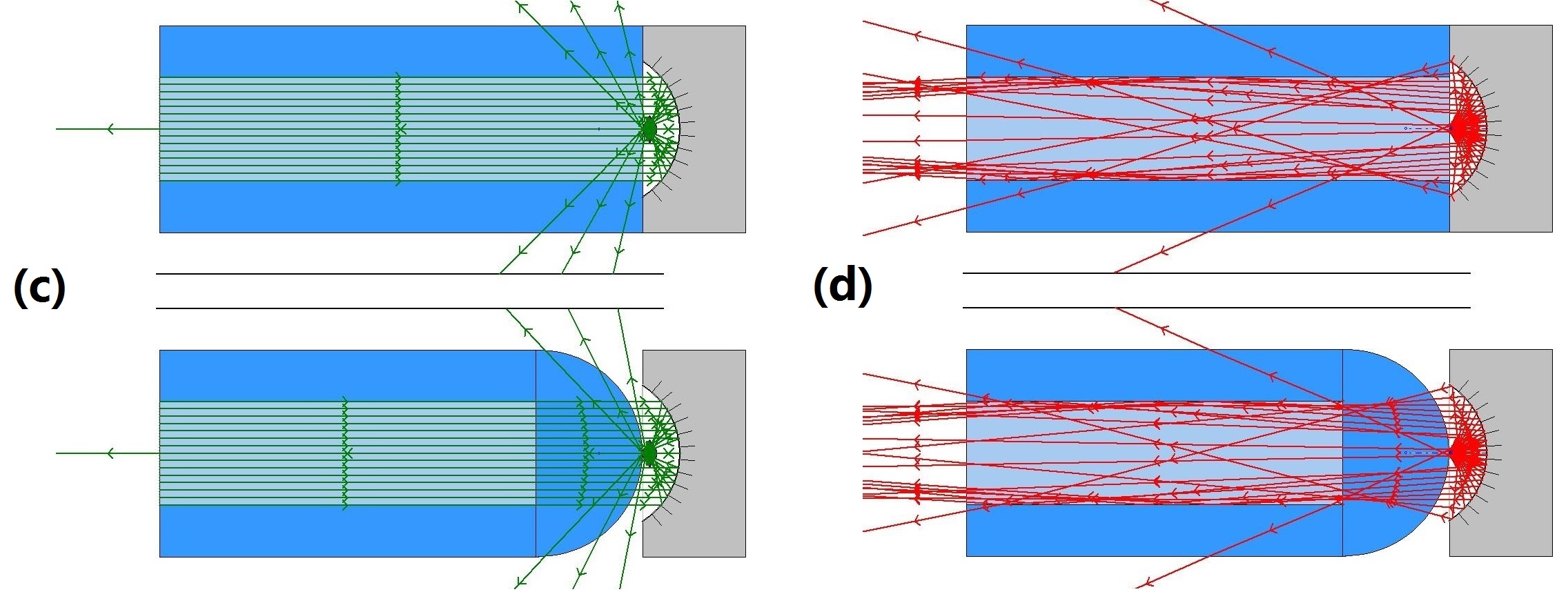}}
\caption{A schematic diagram of the micro-concave (MC) mirror for increasing the NV center (a) excitation and (b) fluorescence collection.  Ray tracing simulations are shown in  (c) and (d).  In the simulation, the mirrors are spherical MC mirrors with a radius of 120 \(\mu \)m; the refractive index of the diamond, optical fiber core and cladding are set to 2.4, 1.47 and 1.45 respectively for both green and red light. The diamond is located at the focal point of the mirror. (The simulations are done by Optgeo (http://jeanmarie.biansan.free.fr/optgeo.html)).}
\label{fig: Figure1}
\end{figure}

The working principle, schematic diagrams, and ray optics fluorescence excitation and collection simulations of the MC mirror-micro-crystal-optical fiber system are shown in figure (Fig.\ref{fig: Figure1}). Though the micro-crystals have high density of NV defects, because of low optical absorption cross-section of the NV defects, except for a small portion of the laser light being absorbed by the dense NV centers, most of the laser light exiting the optical fiber will pass through the diamond. In this case, an MC mirror located with its focal point at the diamond center will reflect and focus the unabsorbed laser light back into the diamond to excite the NV centers. This enhancing the excitation efficiency (Fig.\ref{fig: Figure1} (a) and (c)). Moreover, because of the unstructured surface of the diamond, the excitation laser undergoes partial light-trapping due of total-internal refection at the surface. Hence the excitation laser light undergoes many rounds of reflection within the diamond crystal before being excited from the surface; thankfully, the presence of the MC mirror further couples the laser light into the diamond thus promoting further excitation of NV centers. Simultaneously, the MC mirror will also reflect and focus most of the fluorescence that could exit out of the diamond in the opposite direction of the optical fiber back into the optical fiber within its light acceptance cone. This way the MC-fiber system enhance the fluorescence collection from a NV micro-crystals(Fig.\ref{fig: Figure1} (b) and (d) ).

\begin{figure}[htbp]
\centering
\fbox{\includegraphics[width=0.92\linewidth]{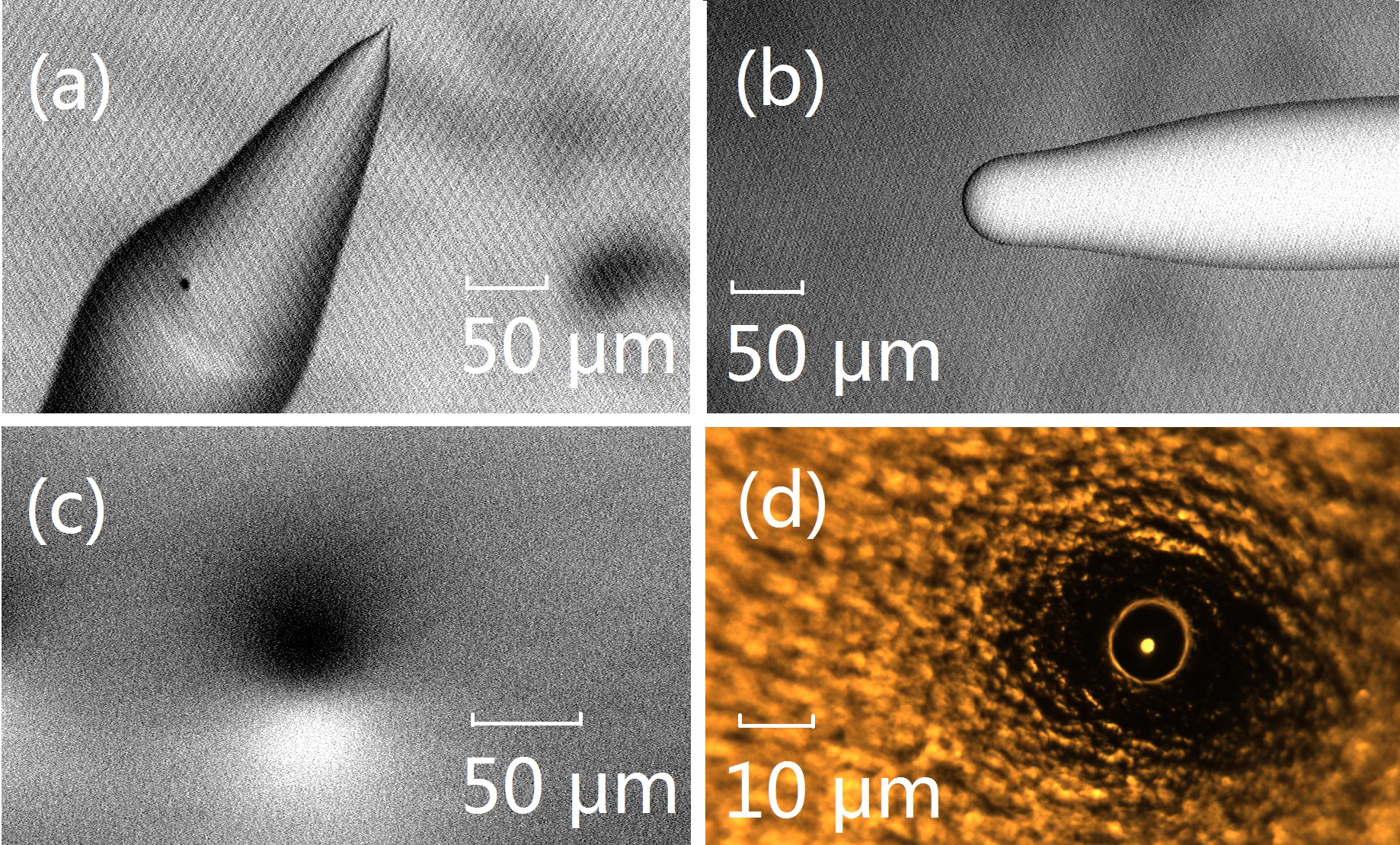}}
\caption{(a) and (b): scanning electron microscope (SEM) image of two prepared tip-sphered optical fibers with different sphere shapes. (c) SEM image of a fabricated MC mirror on aluminum (Al) foil; (d) optical microscopic image of a fabricated aluminum MC mirror focused the illumination light and formed a bright spot in the middle and a bright circle encircling it.}
\label{fig: Figure2}
\end{figure}

The MC mirrors are fabricated by slightly pressing the center of a small section ($ < $2 mm) of thin aluminum (Al) foil ($\sim$15 \(\mu \)m in thickness; purchased from a local supermarket) with a optical fiber with it's tip shaped as a hemispherical end on an ultra-precision fiber optic alignment stage (M-561D-XYZ, Newport, USA). The Al foil is first glued on to a small plastic sheet with a hole at the center. The spherical tipped optical fiber is fabricated by using arc discharge of a tapered optical fiber tip. The desired tapper and the sphering is obtained by heating and simultaneously stretching in an electric fusion splicer (Fitel S153A, Japan). The shape and focal length of the mirror can be adjusted through modifying the fabrication parameters such as the stretching force and the arc discharging current. During the initial phases, we made a set (array) of 5 to 10 MCs and individually tested the enhancements. The enhanced collection, normalized to the maximum of the fluorescent signal deviate less than 1-5 percent in total. Hence for subsequent studies, we just fabricate one MC and that works well as expected. In other words the good-micro mirrors turn-out ratio is near unity taking into consideration the 1-5 $\%$ tolerances. The MC mirror's shape and focal length is quite robust and will not change once the tip-sphered optical fiber is chosen. The fabricated MC mirror is quite stable against drifts and other conditions: even after months it still preserves its shape and gives the same efficiency of enhancements. Fig.\ref{fig: Figure2} (a), (b) and (c) show scanning electron microscope images of two tip-sphered optical fibers and one fabricated Al MC mirror. Fig.\ref{fig: Figure2} (d) shows one fabricated Al MC mirror focused the illumination light and formed a bright spot in an optical microscope image. 

\begin{figure}[htbp]
\centering
\fbox{\includegraphics[width=0.92\linewidth]{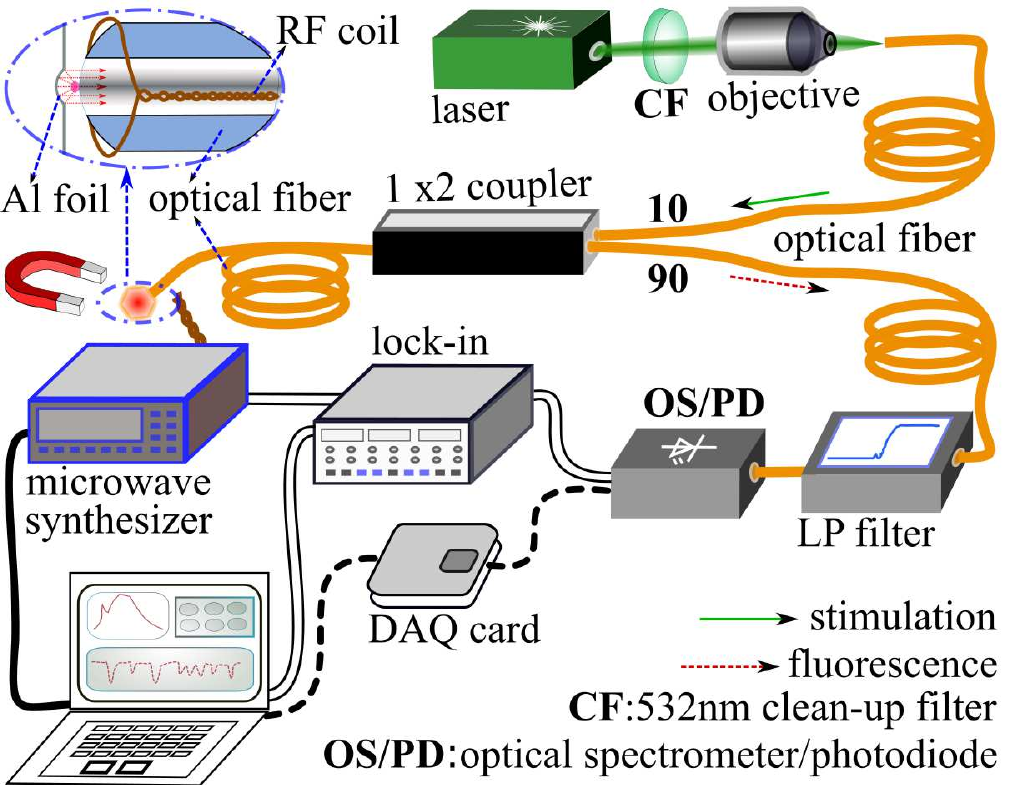}}
\caption{Experimental setup for assessing the MC mirror's NV  fluorescence excitation and collection enhancing. The inset depicts the set of the diamond, the Al MC mirror (Al foil) and the microwave conducting wire (RF coil). In the optically detected magnetic resonance (ODMR) scanning experiments: when no lock-in amplifier is used, the dashed line was connected instead of the three double-solid lines, and the photo-diode was read by the data acquisition (DAQ) card; when the lock-in amplifier is used, the three double-solid lines were connected instead of the two dashed lines, and the photo-diode was read directly by the lock-in amplifier.}
\label{fig: Figure3}
\end{figure}

We used the experimental setup depicted in Fig.\ref{fig: Figure3} to study and quantify the enhancement of the MC mirror-system's NV fluorescence excitation and collection efficiency. The excitation light (solid arrows in  Fig.\ref{fig: Figure3}) from a 532nm laser source is collimated and passed through a clean-up filter, and then coupled into the 10 percent port of a 90:10 $1\times2$ fiber optical coupler ( Beijing XingYuan AoTe technology Co., Ltd, China) using an objective (10$\times$, NA=0.1). About $\sim$10\% of the laser power is guided to the $\sim$8-um diamond glued using an UV curing glue at the center of the sphered end of the graded index multi-mode fiber (GIF625, Thorlab). 

The fluorescence (dashed arrows in Fig.\ref{fig: Figure3}) is collected by the same fiber and guided back into the input port of the fiber optical coupler; $\sim$90\%. The collected fluorescence light passes through a long-pass (LP) filter (cut-off wavelength 615nm) and is detected by an optical spectrometer (BTC655, B\&W tek, USA) or photo-diode (APD430A/M, Thorlab) (OS/PD in Fig.\ref{fig: Figure3}).The micro-size diamond is crushed from NV centers enriched in an HPHT (high pressure,high-temperature) type-Ib single crystal which was electron-irradiated and subsequently annealed. It should be noted that the laser light exiting out of our optical fiber is unpolarized, so the orientation of the diamond crystal will have little effect on the NV excitation efficiency \cite{Acosta2010PRB}. The Al MC mirror is mounted on a precision adjustable three-dimensional translation stage facing the diamond on the sphered fiber end (inset of Fig.\ref{fig: Figure3}). For experiments that require signal acquisition using a lock-in amplifier or using a DAQ card the connections depicted by the dashed line and double-solid lines in Fig.\ref{fig: Figure3} were swapped. 

\begin{figure}[htbp]
\centering
\fbox{\includegraphics[width=0.92\linewidth]{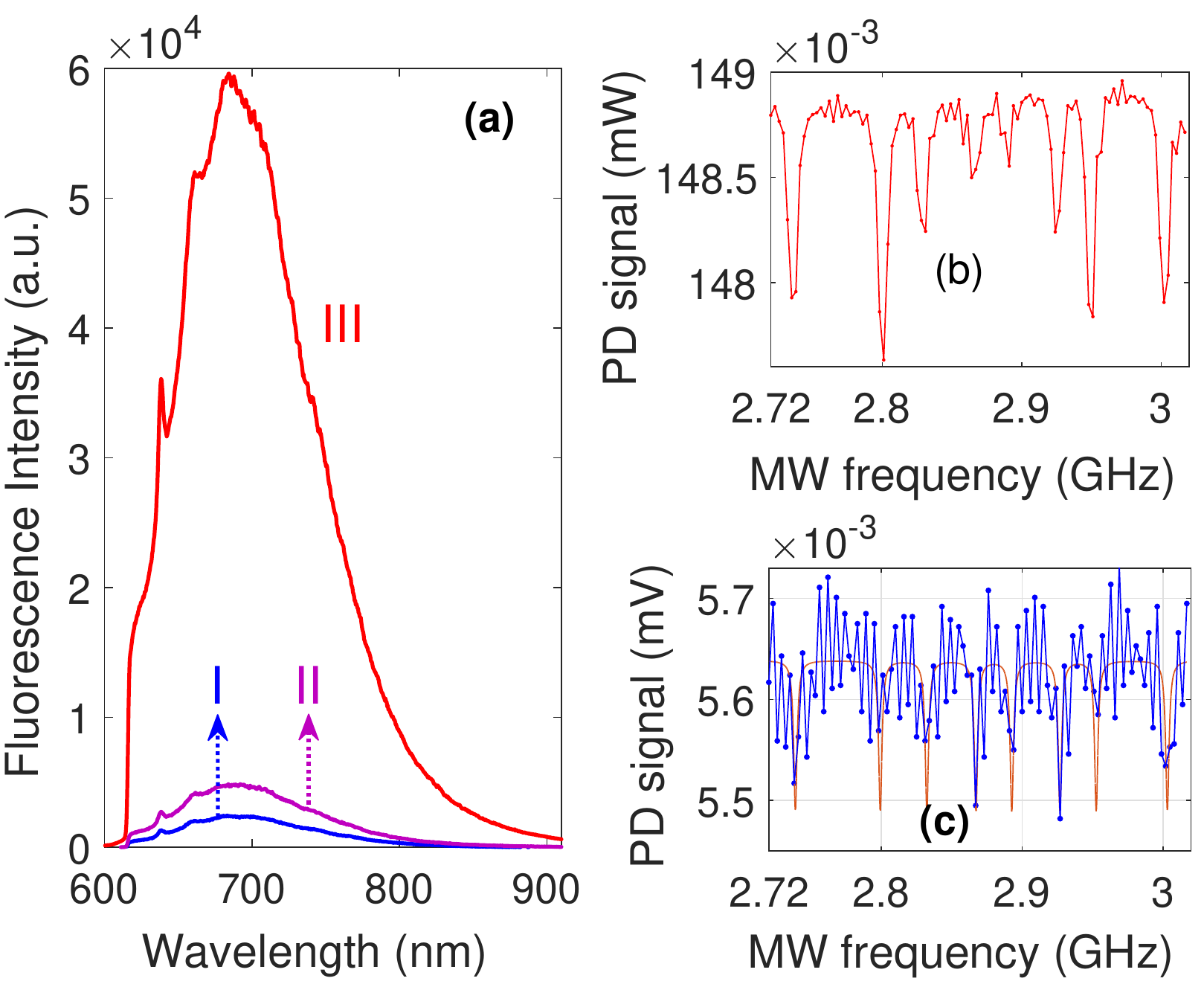}}
\caption{(a) The measured fluorescence spectra of a $\sim$8-\(\mu \)m diamond glued on a sphered optical fiber end in three different configurations: Case I (not using any mirror), Case II (facing an Al flat mirror) and Case III (diamond at the focus of the Al MC mirror). ODMR spectra of the diamond without any averaging (b) case III and (c) Case I. The laser power( $\sim$0.49 mW) and magnetic field ( $\sim$2.74 mT) are kept the same.}
\label{fig: Figure4}
\end{figure}

For systematically studying the fluorescence enhancements, we recorded the spectrum of the diamond placed on the sphered fiber end in three different configurations. Case I- Diamond is placed far away from the MC mirror (not using any mirror), Case II, The diamond is facing a plane mirror (focus at infinity). Case III, where the diamond is approximately located at the focal point of the MC mirror. The results are shown in Fig.\ref{fig: Figure4} (a) and the respective configurations are labeled as cases I, II and III. The incidence laser power (into the fiber) used for these experiments is $\sim$0.49 mW. From the results of Fig.\ref{fig: Figure4} (a) we could observe case III (with MC) shows a considerable increase in the collected fluorescence. To quantify the results, we calculate the integrated area from the obtained spectrum (from case III) and compare that to that of not using any mirror (case I). We get a modest value of over 25 times increase in the fluorescence collection when the MC mirror is placed at the focus (case III) Fig.\ref{fig: Figure4} (a)). 

For studying the advantages of enhanced fluorescence collection on magnetometry applications, we performed the optically detected magnetic resonance (ODMR) on the micro-crystals, fiber, MC system. As an addition to the previous setup, we placed a permanent magnet and a microwave coil (RF coil) near the diamond. The NV defects in a diamond micro-crystals occupy four crystal orientations; hence for these studies we oriented the field arbitrary to the crystal orientation such that we obtain eight ODMR transitions (two for each of the crystal orientations). We could also orient the magnetic field to the NV quantization axis of a crystal orientation to obtain fewer lines and a better contrast. This is done by following the ODMR transition frequencies and ensuring the lines ( $m_{s}=0\rightarrow\pm1$) split to maximum and also symmetrically to that of zero-field-splitting (2870 MHz). For the ODMR experiments, a permanent magnetic field is placed perpendicular to the optical fiber and fine adjustments are made to align the B-field. We recorded the ODMR spectrum shown in Fig.\ref{fig: Figure4} (b) and (c) show the results obtained by sweeping the microwave frequency and synchronously recording the florescence intensity for case I and III. It is evident that in case III we can easily obtain the ODMR spectrum in a single-scan (without averaging) (Fig.\ref{fig: Figure4} (b)); while the results from case I  (Fig.\ref{fig: Figure4} (c)) is very noisy in a single-scan experiment. For a proper comparison, we have fixed the excitation laser power to be the same for both the cases shown in (Fig.\ref{fig: Figure4} (b) and (c)).

\begin{figure}[htbp]
\centering
\fbox{\includegraphics[width=0.92\linewidth]{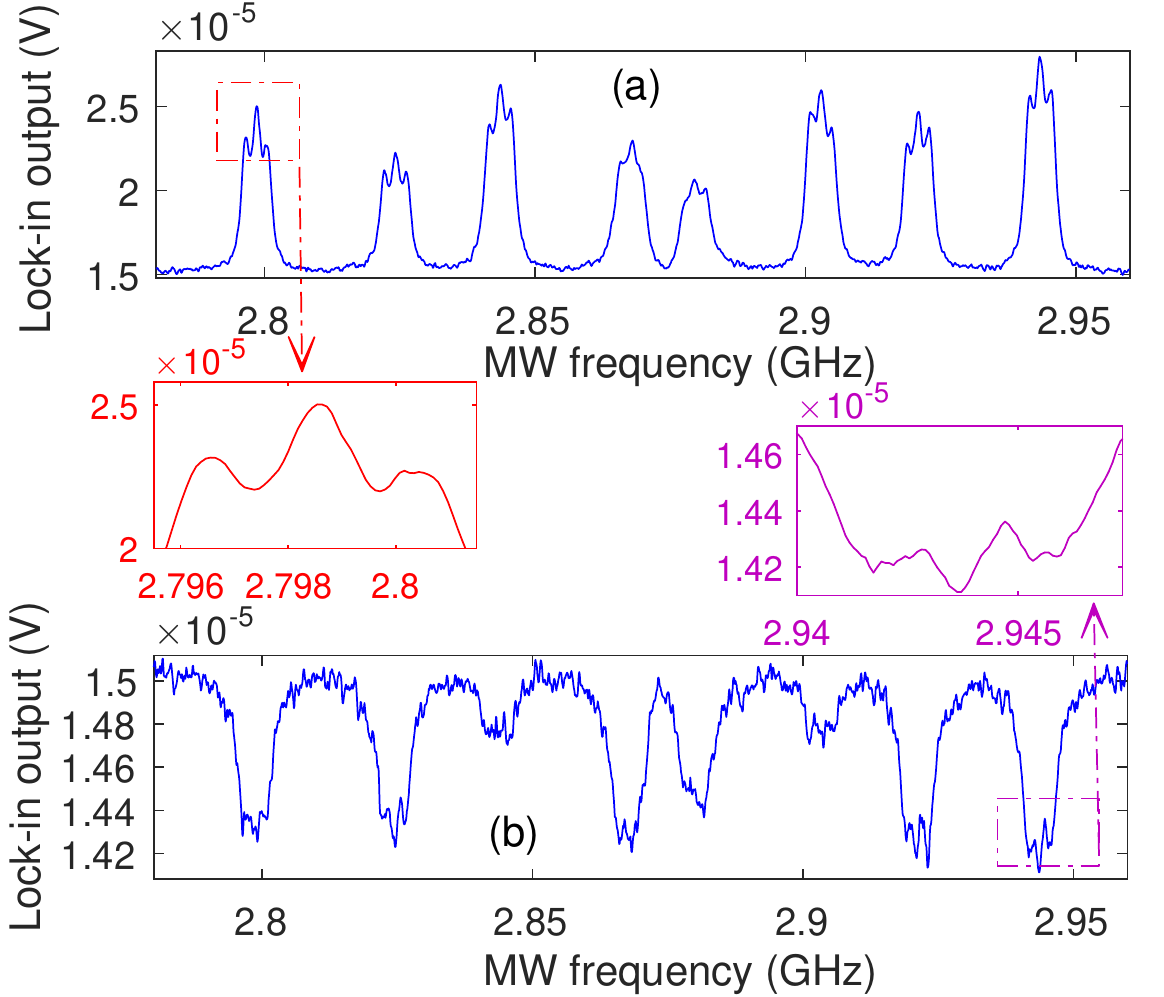}}
\caption{20 times averaged lock-in amplifier outputs of the ODMR spectra of a $\sim$8-\(\mu \)m diamond glued on a sphered optical fiber end (a) located at the focal point of the MC mirror (case III) and (b) facing no mirror (case I). Insets are enlarged spectra of Nitrogen hyperfine splitting of the NV centers.  The laser power( $\sim$ 0.49 mW) and magnetic field ($\sim$1.52 mT) values are the same. }
\label{fig: Figure5}
\end{figure}

To obtain a high resolution ODMR spectrum we have used a lock-in amplifier configuration. The microwave field that drives the NV electron spins is amplitude modulated (at a modulation frequency of 25 KHz and a depth of 100\%) and the signal is recorded using a photo-diode. When the lock-in technique is applied, we get ODMR signals with significantly high SNR. The Nitrogen hyperfine lines are clearly visible (inset in Fig.\ref{fig: Figure5}). Comparing the ODMR results of case III (Fig.\ref{fig: Figure5} (a)) and case I (Fig.\ref{fig: Figure5} (b)) we could certainly ascertain that much better SNR of the ODMR signal is obtained when the diamond is positioned at the focal point of the MC.

\begin{figure}[htbp]
\centering
\fbox{\includegraphics[width=0.92\linewidth]{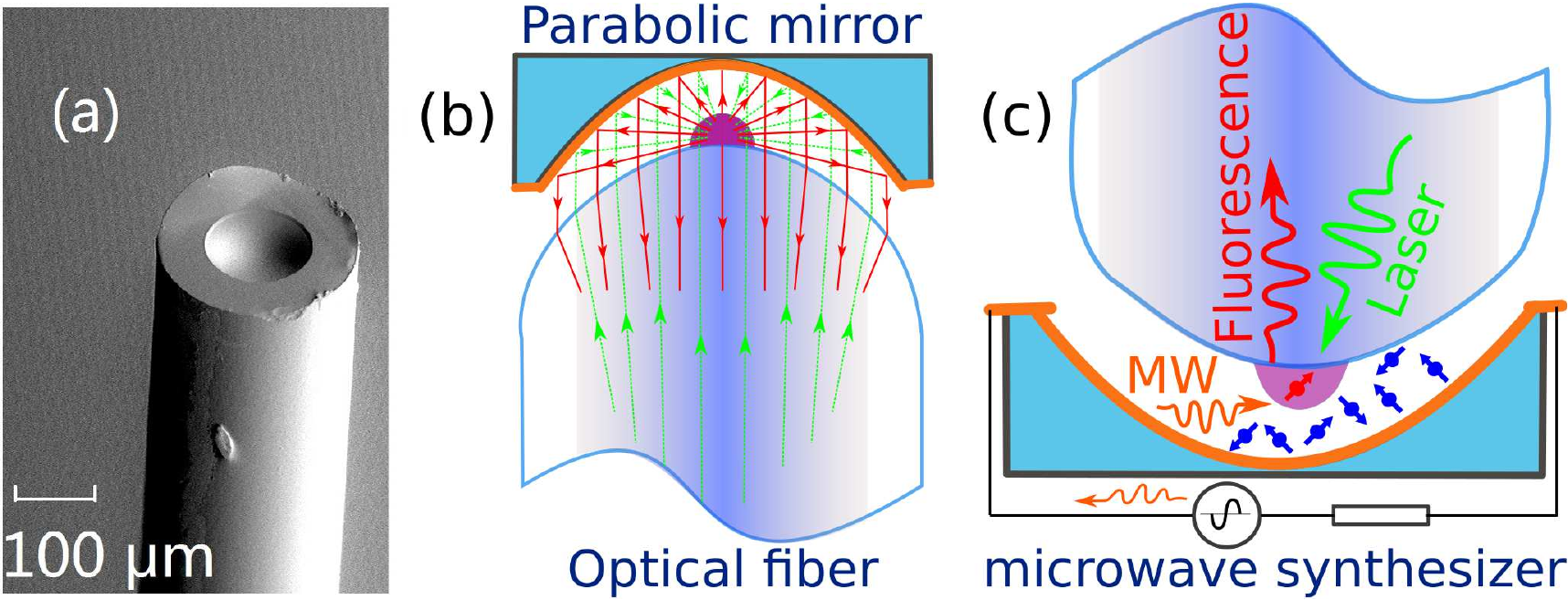}}
\caption{  (a) SEM image of an fabricated MC mirror. (b). Principle schematic diagram of combining the solid immersion lens (SIL) with the parabolic shape MC mirror to further enhance the fluorescence collection efficiency of NV centers. (c). A possible usage setup: the MC mirror is used as a test sample keeper, simultaneously, it is connected to the microwave synthesizer for delivering microwave to the NV centers.}
\label{fig: Figure6}
\end{figure}

 The fabrication of MC mirror using an Al film is simple and straight forward, but its relatively large size reduces the compactness of the optical fiber-based NV sensor systems. To efficiently solve this issue in a simple way, we integrated an MC mirror directly on an optical fiber. The fabrication is done by first etching the graded index fiber GIF625 flat end face with 40\% hydrofluoric acid for about 1 minute. This forms a smooth concave surface aligned to the core of the fiber. The concave etched fiber is then coated with thin film of gold done by thermal deposition. Fig.\ref{fig: Figure6} (a) shows the scanning electron microscopic (SEM) image of one fabricated on-fiber MC mirror.  

One piece of diamond micro-crystal is taken from NV enriched HTHP synthetic mono-crystalline diamond powder and placed at the center of a flat optical fiber end facing an on-fiber integrated MC mirror. We obtained an efficiency enhancement of about 10 times in fluorescence excitation and collection for such a configuration from a $\sim$10\(\mu \)m diamond. Though the integrated MC system shows a modest 10 times increase together with the compactness of the system, the enhancement obtained in this case is certainly not on par with those achieved with the Al foil MC system. This is because the chemical etching has not formed a perfectly matched MC mirrors that enables focusing the laser into the diamond and focusing the fluorescence back into the light acceptance cone of the optical fiber. We believe that through optimizing the etching parameters or using other controllable micro-fabrication techniques such as CO$_{2}$ laser ablation\cite{Hunger2012} and ion etching\cite{Najer2017}, to form optimized shaped on-fiber MC mirrors such as the parabolic reflector on the fiber end, this compact design also can achieve a high-efficiency enhancement.

In this method, the MC mirror can only help in collecting the fluorescence that has exited out of the diamond. As the diamond has a high refractive index ($\sim$2.4), fluorescence from the NV defects undergoes multiple total internal reflections within the diamond micro-crystal. This trapped light in the diamond cannot be collected by the MC mirror. However, methods such as the solid immersion lens (SIL)  \cite{Hadden2010, Marseglia2011, Jamali2014} and the circular Bragg grating \cite{Li2015, Andersen2018} can be used for extracting fluorescence out of the diamond. Combining the MC mirror with these SIL or circular Bragg grating on the diamond itself will effectively enhance the fluorescence collection efficiency.  For example, Fig.\ref{fig: Figure6} (b) shows the working principle schematic diagram of combining a simple parabolic shaped MC mirror and a fluorescence extracting SIL to further enhance the diamond NV center's fluorescence excitation and collection. Fig.\ref{fig: Figure6} (c) depict how our MC mirror based configuration can be potentially used in liquid sample sensing or integrated with a microfluidic system. In addition the metal MC mirror is also found to be able to act as microwave near-field antenna sufficient to perform NV spin manipulation when it is connected to a microwave synthesizer and properly impedance matched.

We have demonstrated a simple method based on an MC mirror to enhance the laser excitation and also fluorescence collection in an optical fiber-based NV sensor system. Using this technique, we achieve 25 times more fluorescence collection from an NV enriched micrometer-sized diamond attached to an optical fiber end when compared to not using the MC mirror. This enhanced detection efficiency also results in ODMR signals with high SNR. This MC mirror + Fiber system method can also be applied to other solid-state quantum emitters such as silicon-vacancy center in diamond, quantum dots, and rare-earth ion-based color centers. Additionally, we made the system very compact by using a $1\times2$ fiber optical coupler to replace the bulky optical elements based optical path splitter in the fiber-based NV quantum sensor system. This single-ended sensor configuration together with the improved SNR could be of immense use in portable and robust magnetic sensor applications.\\

Funding was provided by the Max-Planck Society, Niedersächsisches Ministerium f\"{u}r Wissenschaft und Kultur, DFG Research Centre Nanoscale Microscopy and Molecular Physiology of the Brain, and JSPS KAKENHI (No. 17H02751).


\pagebreak
\widetext
\begin{center}
\textbf{\large appendix}
\end{center}
\setcounter{figure}{0}
\makeatletter
\renewcommand{\thefigure}{Ap\arabic{figure}}
\renewcommand{\citenumfont}[1]{#1}

\begin{figure}[h]
\centering
\includegraphics[width=8cm]{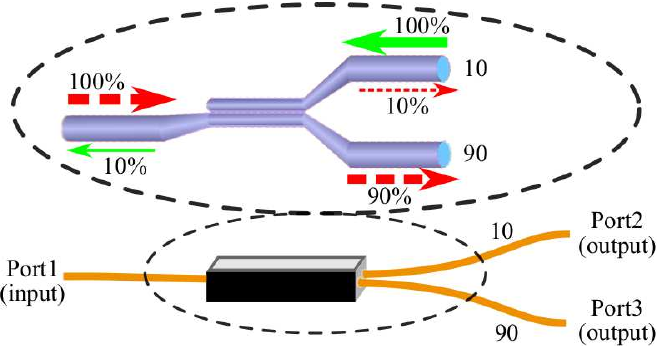}
\caption{Schematic diagram of a 90:10 $1\times2$ fiber optical coupler. Red (dashed) arrows show the fluorescence path and green (solid) arrows show the laser light path in our experiments.}
\label{fig: FigAp1}
\end{figure} 

\begin{figure}[h]
\centering
\includegraphics[width=8cm]{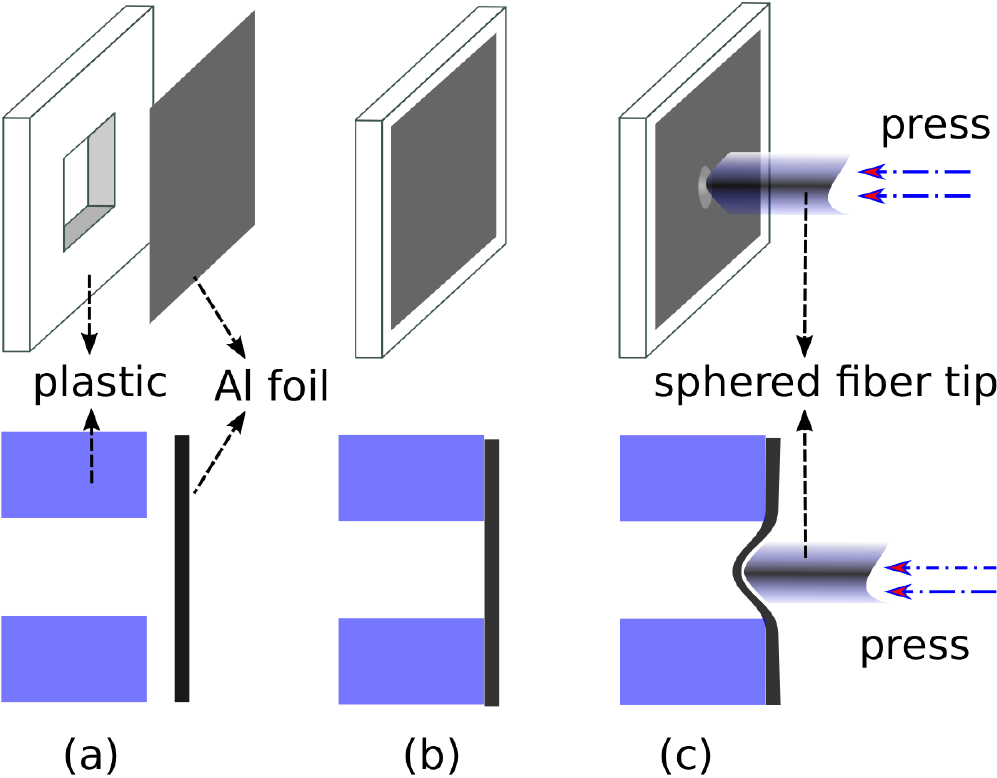}
\caption{The aluminum based micro-concave mirror fabrication procedure: (a) prepare a center-empty thick plastic sheet and cut a section of aluminum foil (Al foil) in size of the plastic sheet; (b) glue the aluminum foil on the plastic sheet; (c) lightly press the middle of the aluminum foil with a tip-sphered optical fiber  to form the micro-concave mirror.On top is a three-dimensional view and at the bottom is the side view.}
\label{fig: FigAp2}
\end{figure}

\end{document}